\documentclass[runningheads]{llncs}

\usepackage{graphicx}
\usepackage{booktabs}
\usepackage{amsmath}
\usepackage{amssymb}
\usepackage{xcolor}
\begin{document}
\title{Pre-text Representation Transfer for Deep Learning with Limited \& Imbalanced Data : Application to CT-based COVID-19 Detection}
%
%

\author{Fouzia Altaf\inst{1}\orcidID{0000-0002-3185-5058} \and
Syed M.S Islam\inst{1}\orcidID{0000-0002-3200-2903} \and
Naeem K. Janjua\inst{1}\orcidID{0000-0003-0483-8196} \and Naveed Akhtar\inst{2}\orcidID{0000-0003-3406-673X}}


%

\institute{Edith Cowan University Joondalup, Australia \and
University of Western Australia Crawley, Australia
}

\maketitle              
\begin{abstract}
Annotating medical images for disease detection is often tedious and expensive. Moreover, the available training samples for a given task  are generally scarce and imbalanced. These conditions are not conducive for learning effective deep neural models. Hence, it is common to `transfer' neural networks trained on natural images to the medical image domain. However, this paradigm  lacks in performance due to the large domain gap between the natural and medical image data. To address that, we propose a novel concept of Pre-text Representation Transfer (PRT). In contrast to the conventional transfer learning, which fine-tunes a source model after replacing its classification layers, PRT retains the original classification layers  and updates the representation layers through an unsupervised pre-text task.    
The task is performed with (original, not synthetic) medical images, without utilizing any annotations. This enables   representation transfer with a large amount of training data.  This high-fidelity representation transfer allows us to use the resulting model as a more effective feature extractor. Moreover, we can also subsequently perform the traditional transfer learning with this model. We devise a collaborative representation based classification layer for the case when we leverage the model as a feature extractor. We fuse the output of this layer with the predictions of a model induced with the traditional transfer learning performed over our pre-text transferred model. The utility of our technique for limited and imbalanced data  classification problem is demonstrated with an extensive five-fold evaluation for three large-scale models, tested for five different class-imbalance ratios for CT based COVID-19 detection. Our results show a consistent gain over the conventional transfer learning with the proposed method.

\end{abstract}

\vspace{-3mm}
\section{Introduction}
\vspace{-2mm}
In the medical imaging domain, data labelling requires medical experts, who must carefully analyse the samples to provide the correct annotation. Not only that this process is tedious, expensive and strongly reliant on the availability of  medical experts, the data itself suffers from plenty of challenges. First, it is common that the positive samples of a disease are much rarer than the negative samples. This naturally creates an imbalance in the data, which is particularly challenging to induce unbiased computational models using that data. Second, for the geographically constrained facilities, both positive and negative samples are often too few to effectively train a computational model that can facilitate automated disease detection.
Incidentally, global data sharing through public repositories also fails to fully resolve these issues due to the data privacy constraints. Whereas medical images are  easily searchable content on the internet, their annotations related to a specific diagnostic task are seldom available. 

It is well-established that deep learning~\cite{lecun2015deep} can induce  computational models that can achieve expert-level accuracy for many disease detection tasks using medical images~\cite{esteva2021deep}. This fact has led to a wave of deploying deep learning solutions in medical image analysis~\cite{altaf2019going}. However, this technology can only perform effective computational modelling if it is provided with a large amount of training data (e.g.,~a million samples). For the medical tasks, these samples  need to be appropriately annotated by the experts. Thus, the challenges noted in the preceding paragraph present a bottleneck for fully exploiting deep learning in medical image analysis. Currently, Transfer Learning (TL)~\cite{yu2022transfer} is a common strategy to side-step this bottleneck~\cite{altaf2021boosting}, \cite{altaf2021novel}, \cite{roberts2021common}.

Transfer Learning takes a deep learning model pre-trained for a \textit{source} domain, and fine-tunes it with a \textit{target} domain data. For the medical tasks, natural images usually form the source domain~\cite{altaf2019going} due to their convenient annotations.  The central idea behind TL is that by using a large amount of training images, the pre-trained model (a.k.a.~source model) learns a detailed representation of the source domain. This representation also encodes the primitive patterns that form the fundamental image ingredients. Since the target medical domain also comprises images, it is likely that a slight modification to this encoding can already  be sufficient to represent the target domain samples reasonably well. Transfer learning seeks to induce the desired  modification with the scarcely available data for the medical task at hand. 

Altaf et al.~\cite{altaf2021boosting} recently noted that the large domain gap between the natural and medical images compromises the performance of TL for the medical tasks. They argued that this large gap requires proportionally large data of the target domain for an effective model transfer. Hence, they proposed to first transfer the source model to the target domain with a large-scale annotated medical data, albeit under a different auxiliary imaging modality.
Their assumption is that, for a target data modality (e.g.,~CT scans), large-scale annotated samples are available for a related auxiliary modality  (e.g.,~radiographs) in the medical domain.
First transferring the model to the medical domain with the auxiliary modality, and then transferring it further to the target modality, is shown to improve the TL performance~\cite{altaf2021boosting}. Though effective, availability of annotated auxiliary large-scale  data is still a strong assumption for the medical imaging domain. Moreover, the imaging modality disparity within the target domain (e.g.,~CT scans vs radiographs) can still be problematic.       

This paper introduces a novel concept of Pre-text Representation Transfer (PRT), which enables effectively transferring the source domain representation to the target domain without any data modality disparity, or assuming additional data annotations. It formulates an unsupervised learning task, termed \textit{pre-text} task inspired by the self-supervised learning literature~\cite{misra2020self}, which enables the use of a large amount of original target domain data. This data is un-annotated or has irrelevant annotations w.r.t.~the target downstream task. We meticulously transfer the source domain `representation' to the medical domain using this task. This allows us to use the transferred representation both as an effective feature extractor and as a source model to perform further transfer learning. We leverage both options, and fuse their predictions to compute the final output. In the process, we also adapt a collaborative representation  scheme to serve as a classification layer for our  feature extractor, such that its predictions can be intelligibly fused with the predictions of the transferred model for a performance boost.  

Owing to the sensitivity of  Computed Tomography (CT) to COVID-19~\cite{wynants2020prediction}, and available benchmark studies on the exploration of transfer learning for the CT-based COVID-19 detection~\cite{altaf2021resetting}, we showcase the efficacy of our approach for this problem with an extensive evaluation that performs over 75 deep learning model training sessions. With a five-fold validation for three different models, using five different imbalance ratios of the limited CT training images, we demonstrate a consistent improvement over the conventional transfer learning performance with our technique. 
The main contributions of this paper can be summarized as follows.
\begin{itemize}
    \item We introduce a novel concept of Pre-text Representation Transfer (PRT) that allows effective transfer of the representation component of a model to the target domain on (large) unlabelled data.
    \item We develop a method that leverages the  model resulting from PRT as a feature extractor, and fuses its predictions with  subsequent transfer learning over the resulting model. 
    \item With extensive five-fold experiments for CT-based COVID-19 detection, we establish the effectiveness of our method for limited and imbalanced data.
\end{itemize}

\begin{figure}[t]
    \centering
   \includegraphics[width = 0.99\textwidth]{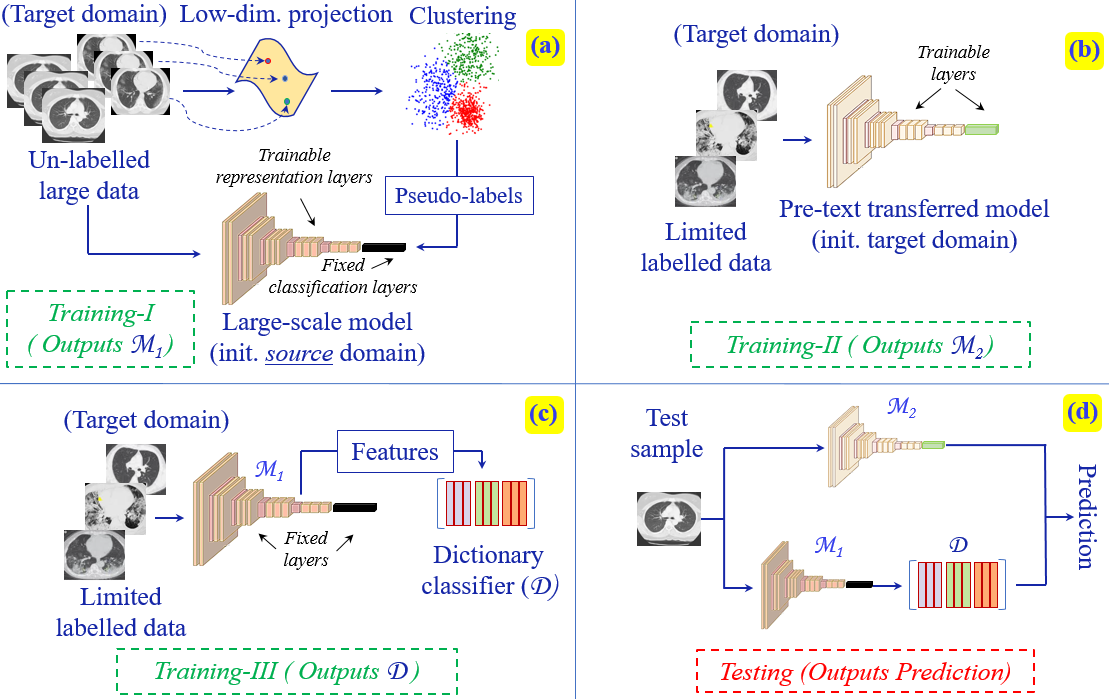}
    \caption{\textbf{(a)} A large amount of un-annotated data of the target domain is clustered in a lower dimensional space to compute pseudo-labels. Using these labels, pre-text representation transfer is performed using a source domain model. \textbf{(b)} The pre-text transferred model $\mathcal M_1$ is further modified with the conventional Transfer Learning to induce model $\mathcal M_2$. \textbf{(c)} The $\mathcal M_1$ is used as a feature extractor to construct a feature dictionary $\mathcal D$ of the labelled training data. \textbf{(d)} A classifier defined by $\mathcal D$ fuses its predictions with the output of $\mathcal M_2$ to provide the final prediction.}
    \label{fig:main}

\end{figure}

\section{Proposed method}

We illustrate the central concept of our Pre-text Representation  Transfer (PRT) based method in Fig.~\ref{fig:main}. To concisely  present our contribution, we first formalize the model transfer mechanism under the traditional transfer learning (\S~\ref{sec:TIII}). Then, the components of our methods are discussed in detail (\S~\ref{sec:TI}-\ref{sec:test}). 

\vspace{-3mm}
\subsection{Traditional model transfer}
\label{sec:TMT}
From a computational perspective, we typically see medical image based disease detection as an image classification task. Let $\mathcal{M}({\bf x}\sim\mathcal X, \boldsymbol{\Theta}): {\bf x} \rightarrow \boldsymbol\ell_x$ denote an image classifier, where ${\bf x}$ is sampled from  the image distribution $\mathcal X$, $\boldsymbol{\Theta}$ denotes the model parameters, and $\boldsymbol\ell_x \in [0,1]^L$ is an $L$-dimensional binary class vector, encoding the labels of $L$ classes. For a deep neural model, a large amount of annotated samples is normally required to induce an effective $\mathcal M(.,.)$. Owing to the unavailability of adequate samples, medical image classification often  seeks a transform $\boldsymbol\Psi: \widetilde{\mathcal{M}}({\bf z} \sim \mathcal Z, \widetilde{\boldsymbol{\Theta}}) \rightarrow \mathcal{M}({\bf x}\sim \mathcal X, \boldsymbol{\Theta})$, where $\mathcal Z$ is the distribution over natural images, and $\widetilde{\mathcal M}(.)$ is a natural image classifier with parameters $\widetilde{\boldsymbol{\Theta}}$. For ${\bf z}$, normally $\boldsymbol\ell_z \in [0,1]^{\widetilde{L}}$, s.t.~$\widetilde{L}>> L$.
The function $\boldsymbol \Psi$ can be understood as the core model transfer function in the traditional transfer learning. 
Due to the hierarchical nature of the neural models, we can write  $\widetilde{\mathcal{M}} (.,.) =  \widetilde{\mathcal C}(\widetilde{\boldsymbol{\Theta}_C}, \widetilde{\mathcal R} (\widetilde{\boldsymbol{\Theta}_R}, .))$. Here, $\widetilde{\mathcal C}(.,.)$ is composed of the deeper `classification' layers of the model, having their respective parameters $\widetilde{\boldsymbol{\Theta}_C}$. The remaining earlier layers with  parameters $\widetilde{\boldsymbol{\Theta}_R}$ are  given by $\widetilde{\mathcal R}(.,.)$, which encode a  `representation' of the distribution $\mathcal Z$ of the source domain. 

In the traditional transfer learning, the function $\boldsymbol\Psi$  replaces $\widetilde{\boldsymbol\Theta_C}$ with a new set of parameters $\boldsymbol\Theta_C$, and provides a slightly modified version of $\widetilde{\boldsymbol{\Theta}_R}$, say $\widehat{\boldsymbol{\Theta}_R}$ for the representation of the target image distribution. The eventual transferred model for the medical domain then becomes $\mathcal M(.,.) =  {\mathcal C}(\boldsymbol{\Theta}_C, (\widehat{\mathcal R} (\widehat{\boldsymbol{\Theta}_R}, .))$. Both  $\boldsymbol{\Theta}_C$ and $\widehat{\boldsymbol{\Theta}_R}$ are learned with the samples of $\mathcal X$. However, due to the limited data availability, the latter is only a slight modification of $\widetilde{\boldsymbol{\Theta}_R}$ under an extremely small learning rate. It is known that if the domain gap $|| \mathcal{Z} - \mathcal X||$ is large, only a slight perturbation to the source domain representation  may not be sufficient to implement $\boldsymbol\Psi$~\cite{altaf2021boosting}. This is a major  limitation of the traditional transfer learning, which renders the model  transfer between the domains with a large covariate shift ineffective.

\subsection{Pre-text representation transfer (Training-I)}
\label{sec:TI}
Within the mainstream Machine Learning literature, self-supervised representation learning~\cite{ericsson2021self} is an established paradigm to address the limited training data problem. A stream of methods following this paradigm, uses the notion of a \textit{pre-text} task~\cite{albelwi2022survey}, which does not require human labelling of the data. For instance, differentiating between known transformations of a given image and other images, is a pre-text task used by the contrastive learning methods~\cite{zhang2020contrastive}. The hope here is that by solving the pre-text task with a very large amount of unlabelled data, the model can still learn the general representation of the domain. This representation can then be fined-tuned to a given task using a limited amount of labelled data under the optimization objective of that downstream task. 

We highlight two challenges in directly using the pre-text task based self-supervised learning for the medical domain. Firstly, it needs to train the original model $\widetilde{\mathcal{M}}(.,.)$ on an enormous amount of data for a very long duration. This extra-ordinary computational requirement is a major limitation of the self-supervised learning paradigm in general. This issue becomes even more pronounced in the sub-domain of medical imaging due to resource limitations in  geographically constrained facilities. Secondly, this type of learning does not consider domain transfer for the downstream task. Notice in the preceding paragraph that the pre-text and the downstream tasks are both defined in the same domain. This  preempts the possibility of transferring a self-supervised model learned with a natural image pre-text task to the medical domain. Addressing  the issues, we introduce the notion of Pre-text Representation Transfer (PRT). 

We let our source model $\widetilde{\mathcal M}({\bf z} \sim \mathcal Z,.): {\bf z} \rightarrow \boldsymbol\ell_z$ to be a large-scale pre-trained model of the natural images.  We then form a set $\widehat{X}$, such that  each element of this set $\widehat{x} \in \widehat{X}$ is still taken from the medical image distribution $\mathcal X$. We further impose that the imaging modality of $\widehat{x}$ is also the same as the modality of our eventual downstream task.
That is, if the downstream task uses chest CT scans, $\widehat{X}$ only contains chest CT images. We are not concerned with the labels of  $\widehat{X}$. Hence, any available image that satisfies our constraint of having the same imaging modality, belongs to $\widehat{X}$. In this work, we scrap the data from  public repositories over the internet to create this set. This set is subsequently transformed into $\widetilde{L}$ clusters. It is emphasized that in our settings,  $\widetilde{L}$ is also the dimensionality of $\boldsymbol\ell_z$. This is intentional, as it helps us in systematically leveraging clustering, which is an unsupervised  process. Our aim here is to transform the representation component $\widetilde{\mathcal R}(\widetilde{\boldsymbol\Theta_R},.)$  of $\widetilde{\mathcal{M}} (.,.)$, such that it can  perform a conservative clustering of $\widehat{X}$ as a pre-text task. To make the process computationally efficient, we perform clustering over  low-dimensional projections of $\widehat{x}$. In this work, we use ResNet50~\cite{he2016deep} features of $\widehat{x}$ for the projection purpose. However, using any other projection method is also viable under the proposed pipeline.      

Using the indices of the resulting clusters as pseudo-labels for $\widehat{X}$, we perform model training for our pre-text representation transfer. To that end, we initialize our network with the natural image model $\widetilde{\mathcal{M}}(.,.)$, and keep the classifier component $\widetilde{\mathcal C}(\widetilde{\boldsymbol\Theta_C}, .)$ fixed during the training. This induces maximal modifications in the representation component $\widetilde{\mathcal R}(\widetilde{\boldsymbol\Theta_R}, .)$ according to $\widehat{X}$. This is in sharp contrast to the traditional transfer learning  that enforces minimal change on the representation component. The model resulting from this training is $\mathcal M_1 ({\widehat{x}} \sim \widehat{X}, \widehat{\boldsymbol\Theta}): \widehat{x} \rightarrow \boldsymbol\ell_{\widehat{x}} \in [0,1]^{\widetilde L}$. In the text to follow, we denote this model  as $\mathcal M_1(.,.)$ for brevity.

\subsection{Subsequent transfer learning (Training-II)}
\label{sec:TII}
Following our notational convention, $\mathcal{M}({\bf x}\sim\mathcal X, \boldsymbol{\Theta}): {\bf x} \rightarrow \boldsymbol\ell_x$ is a model that  results from the traditional transfer learning under  $\boldsymbol\Psi: \widetilde{\mathcal M} (.,.) \rightarrow \mathcal M(.,.)$. It is possible to simply substitute $\widetilde{\mathcal M} (.,.)$ with $\mathcal M_1(.,.)$ in this transformation, where $\mathcal M_1(.,.)$ is the model resulting from \S~\ref{sec:TI}. This can be done  because we have ensured that $\boldsymbol\ell_z, \boldsymbol\ell_{\widehat{x}} \in [0,1]^{\widetilde L}$. In other words, we have not changed the number of labels for the underlying classification task. Hence, $\mathcal M_1(.,.)$ is directly substitutable in the training process.  It can be expected that this substitution can benefit $\boldsymbol\Psi$ because the representation of $\mathcal M_1(.,.)$ is not only in the target domain $\mathcal X$, but it is also strictly restricted to the imaging modality of the samples in $\mathcal X$. Hence, we can subsequently perform the traditional transfer learning on $\mathcal M_1(.,.)$ to obtain a model $\mathcal M_2(.,.)$. This model  modifies the representation component of $\mathcal M_1(.,.)$ only slightly and its classifier component more aggressively, with $10\times$ learning rate. This is in-line with the conventional transfer learning paradigm.

\subsection{Feature extraction and dictionary classifier (Training-III)}
\label{sec:TIII}
The traditional transfer learning updates the whole model, including its representation component. Since the proposed PRT already brings that component of $\mathcal M_1(.,.)$ in the target domain, we can also exploit $\mathcal M_1(.,.)$ separately as an effective feature extractor for the samples in $\mathcal X$. We  leverage this fact in our method. Following \cite{altaf2021boosting}, we use a collaborative representation \cite{akhtar2017efficient} based classifier to predict the labels of the extracted features. The classifier constructs a dictionary $\mathcal D$ with the $\mathcal M_1(.,.)$ features of the training data. This dictionary is used to compute a class probability of a test feature extracted from $\mathcal M_1(.,.)$. The construction of the dictionary is relatively simple. Say, we have `$n$' training samples available for a given class. We first construct a sub-matrix $\boldsymbol d_c = [\mathcal{R}_1(\hat{x}_1), \mathcal{R}_1(\hat{x}_2),..., \mathcal{R}_1(\hat{x}_n)]$ for each class, where $\mathcal R_1(.)$ is the representation component of $\mathcal M_1(.)$. Then, we concatenate the sub-matrices for all the $C$ classes to form $\mathcal D = [\boldsymbol d_1, \boldsymbol d_2,...\boldsymbol d_C]$. This allows us to form a structured dictionary without requiring balanced training data. That is, we allow $\text{col}(\boldsymbol d_i) \neq \text{col}(\boldsymbol d_j )$, s.t.~$i\neq j$ and col(.) computes the number of columns of the matrix in its argument.  
This construction of a structured dictionary is largely inspired by \cite{altaf2021boosting}. However, we extend the prediction mechanism of \cite{altaf2021boosting} to better adapt to our methodology. This is discussed in  \S~\ref{sec:test}.

\subsection{Output predictions (Testing)}
\label{sec:test}

From the implementation perspective, the output of an $L$-class  classifier is a probability vector $\boldsymbol\rho \in \mathbb R^L$, s.t.~$\sum\limits_{i = 1}^L \rho_i = 1$, where $\rho_i$ is the $i^{\text{th}}$ coefficient of $\boldsymbol\rho$. The label `$\ell$' of a sample is predicted as $\ell = \underset{i}{\operatorname{argmax}} <\rho_i>$. In an analogous manner, the existing dictionary based classifier \cite{altaf2021boosting} also chooses to maximize the coefficients of a  vector $\boldsymbol q \in \mathbb R^L$ to predict the class label.  However, there is no external constraint over the coefficients $q_i$ of $\boldsymbol q$. This is problematic because we eventually want to combine the predictions of dictionary based classification with the predictions of a deep model.
Hence, we introduce a constraint in the mechanism, i.e.~$\sum\limits_{i = 1}^L q_i = 1$ to render $\boldsymbol q$ into a probability vector.
This allows a meaningful fusion of $\boldsymbol\rho$ and $\boldsymbol q$ as two probability vectors. We eventually predict the class label for a test image  as $\ell = \underset{i}{\operatorname{argmax}}  <\mathbb E[\rho_i, q_i]>$.

It is noteworthy that besides being tailored for limited training data, our overall technique is intrinsically amenable to data imbalance. This is because, firstly, the central idea of PRT does not assume uniform data clustering. Thereby, it learns a representation without asserting a uniform class distribution in the target domain. Secondly, the used dictionary based classifier~\cite{altaf2021boosting} performs linear operations using an over-complete basis representation~\cite{tovsic2011dictionary}. Provided the availability of relevant basis vectors of the desired class in the dictionary, this scheme does not favor a class because of the relative number of the training samples. In our overall framework,  transfer learning  (\textit{Training II} in Fig.~\ref{fig:main}) is the only  process that conventionally operates under balanced data assumption. However, we perform transfer learning on a model resulting from PRT. Moreover, the output of this model is further fused with the dictionary based classifier's prediction. This compensates for the implicit data balance assumption.

\section{Experiments}
\label{sec:Exp}
The problem of dealing with limited and imbalanced data is particularly relevant  to the medical imaging domain. Considering the currently prevalent  COVID-19 pandemic, we use the task of COVID-19 detection as a test bed for our technique. It is now established that  Computed Tomography (CT) demonstrates even higher sensitivity to COVID-19 than Reverse Transcription Polymerase Chain Reaction (RT-
PCR)~\cite{wynants2020prediction}, \cite{long2020diagnosis}, \cite{fang2020sensitivity}. Hence, advancing our understanding of CT-based COVID-19 detection is particularly important. Thus, in our experiments, we focus on COVID-19 detection using CT images.

\vspace{1.5mm}
\noindent\textbf{\textit{Datasets and settings:} } Our experiments utilize two sets of data. (1) Un-labelled large image set $\widehat{X}$. (2) Labelled limited data. The former is required for  \textit{Training I} in Fig.~\ref{fig:main}. We scrap $10$K samples from the internet repositories ~\cite{yan2018deeplesion}, ~\cite{reeves2009public} for form the set $\widehat{X}$. It is emphasized that we do not require any annotations for $\widehat{X}$, hence the labels provided by the repositories are not used. This is a major advantage of our technique, as it allows us to use any CT image (annotated or not) to improve performance over the downstream task of COVID-19 detection. For (2), we choose the Covid CT Dataset (CCD)~\cite{zhao2020covid} that contains 349 images of infected and 397  images of non-infected patients. A further 48 images from the negative samples were dropped to emulate balanced data. Following \cite{altaf2021resetting}, we employ a five-fold evaluation protocol that sequentially splits the data into training and testing sets. To emulated imbalanced scenarios, we keep 10, 25, 50 and 75\% data points from the positive samples in the training folds (discarding the rest), while the negative samples are always fixed to 349. It is worth emphasizing that these settings are particularly challenging because of the high level of data scarcity and imbalance. We transfer ImageNet models ResNet101~\cite{he2016deep},  VGG16~\cite{simonyan2014very} and DenseNet201~\cite{huang2017densely} in our experiments. These are commonly used standard large-scale models trained over 1 million natural images of 1,000 categories of daily-life objects.  

 \begin{figure}[t!]
    \centering
    \includegraphics[width = 0.9\textwidth]{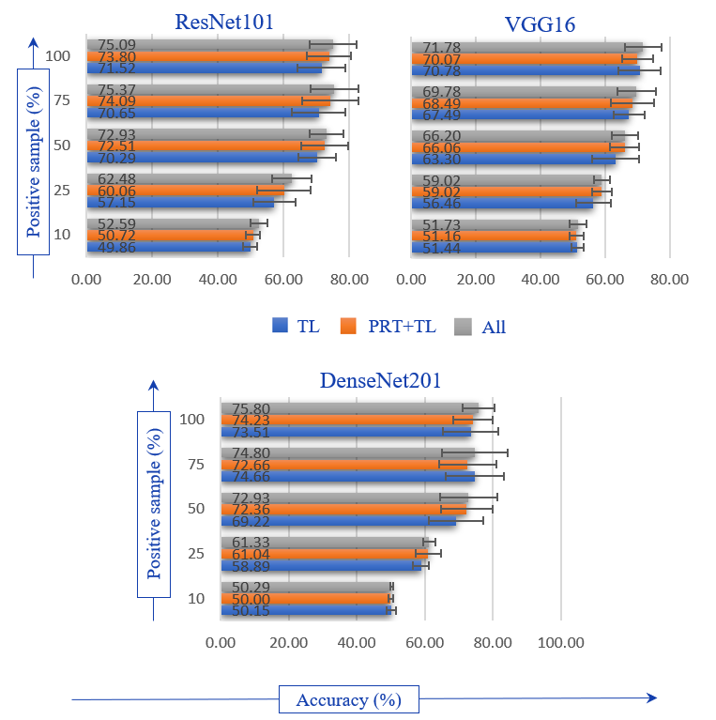}
    \caption{Average accuracies for the five-fold experiments with three models. Results for \textbf{TL:} Transfer Learning, \textbf{PRT+TL:} Pre-text Representation Transfer (PRT) and TL and \textbf{All:} PRT+TL and dictionary based classification. Percentage of COVID-19 positive samples in the training data varies along the y-axis.}
    \label{fig:acc}
\end{figure}

\vspace{1.5mm}
\noindent\textbf{\textit{Implementation details:}} We use the ImageNet models \cite{simonyan2014very}, \cite{he2016deep}, \cite{huang2017densely} provided by Mathworks$^{\tiny\textcircled{c}}$ and conduct experiments with Matlab$^{\tiny\textcircled{R}}$ on an NVIDIA GeForce GTX 1070 GPU with 8GB RAM. Based on our models and setup, $\widetilde L = 1,000$ and $L = 2$. In the \textit{Training I} session - see Fig.~\ref{fig:main}, K-Means clustering is performed with $K = \widetilde L= 1,000$. The PRT is conducted with 15 epochs of training with a learning rate of 3e-4, using a batch size 16. The \textit{Training II} session followed a similar settings, except that we reduced the number of epochs to 7. This is inline with \cite{pham2020comprehensive}. The hyper-parameter settings for \textit{Training III} session followed \cite{altaf2021boosting} for the dictionary based classifier.

\vspace{1.5mm}
\noindent\textbf{\textit{Results:}}
We report the results of our experiments in Fig.~\ref{fig:acc} and Tables~1, 2, and 3. In the figure, we plot the Accuracies (ACC) of the three models, whereas the tables summarise the Specificity (SPE), Sensitivity (SEN) and F1-Score (F1) values of our five-fold experiments. 
Let us denote the true positive outputs as TP, true negatives as TN, false positives as FP and false negatives as FN. The definitions of these metrics can then be given as
\begin{align*}
    \text{SPE} = \frac{\text{TN}}{\text{TN} + \text{FP}} \times 100\% 
    ~,~\text{SEN} = \frac{\text{TP}}{\text{TP + FN}} \times 100\%,
    \end{align*}
    \begin{align*}
    \text{F1} = 2 \times \frac{\text{PPV} \times \text{TPR}} {\text{PPV} + \text{TPR}},
\end{align*}
where we compute PPV as  TP/(TP+FP) and TPR as TP/(TP+FN). We compute Accuracy (ACC) as
\begin{align*}
    \text{ACC} = \frac{\text{TP} +\text{TN}}{\text{TP} +\text{TN} + \text{FP} + \text{FN}} \times 100\%.
\end{align*}

The other abbreviations used in reporting the results have the  following meanings. \textbf{TL:} only conventional Transfer Learning (TL) is used. \textbf{PRT+TL:} Conventional TL is performed over the proposed Pre-text Representation Transfer (PRT). \textbf{All:} The PRT+TL predictions are also fused with the dictionary based classifier predictions.

From the plots in Fig.~\ref{fig:acc}, we can make a few interesting observations. First, there is almost a consistent gain in the accuracy over the conventional TL with our eventual method (i.e., All). Second, in general, both PRT and dictionary based components are able to contribute to the final performance. This verifies our intuition that whereas a subsequent TL on a pre-text transferred model is useful, that model already learns an effective representation of the target domain that can be used for feature extraction. With respect to the TL, the average absolute gains in the accuracy of ResNet101, VGG16 and DenseNet201 for PRT+TL are 2.35, 1.07 and 0.77 respectively. Whereas the overall gains are 3.8, 1.81 and 1.75.  We can also observe analogous   trends in the sensitivity (SEN), specificity (SPE) and F1-score (F1) values for our five-fold experiments in Tables~\ref{tab:ResNet} to \ref{tab:DenseNet}. In general, results corresponding to `All' are the best, followed by PRT+TL, followed by the conventional TL.

\begin{table}[t!]
{\scriptsize 
  \centering
  \setlength{\tabcolsep}{0.1em}
  \caption{Five-fold results for ResNet101. Sensitivity (SEN), Specificity (SPE) and F1-score (F1) score reported for different percentage (\%) of COVID-19 positive samples in  training data. Results provided for \textbf{TL:} Transfer learning, \textbf{PRT+TL:} Pre-text Representation Transfer (PRT) and TL, \textbf{All:} PRT+TL and dictionary based classification. }
    \begin{tabular}{c|c|c|c|c|c|c|c|c|c}
    \toprule
    \textbf{\%} & \multicolumn{3}{c|}{\textbf{TL}} & \multicolumn{3}{c|}{\textbf{PRT+TL}} & \multicolumn{3}{c}{\textbf{All}} \\
\cmidrule{2-10}          & \textbf{SEN} & \textbf{SPE} & \textbf{F1} & \textbf{SEN} & \textbf{SPE} & \textbf{F1} & \textbf{SEN} & \textbf{SPE} & \textbf{F1} \\ \hline
    \midrule
    10    & 97.1$\pm4.1$ & 2.6$\pm1.8$  & 0.66$\pm0.02$  & 97.4$\pm3.4$ & 4.0$\pm2.8$  & 0.66$\pm0.02$  & 99.1$\pm1.2$ & 6.1$\pm6.1$  & 0.68$\pm0.01$ \\ \hline
    25    & 90.7$\pm10.8$& 23.5$\pm13.6$ & 0.68$\pm0.05$  & 85.7$\pm7.9$ & 34.4$\pm17.4$ & 0.68$\pm0.05$  & 90.3$\pm3.5$ & 34.7$\pm13.7$ & 0.71$\pm0.03$ \\ \hline
    50    & 76.2$\pm4.5$ & 64.3$\pm9.8$ & 0.72$\pm0.04$  & 77.3$\pm14.1$ & 67.6$\pm7.1$ & 0.73$\pm0.08$  & 81.3$\pm7.3$ & 64.4$\pm5.7$ & 0.75$\pm0.05$ \\ \hline
    75    & 75.1$\pm16.6$& 66.2$\pm14.9$ & 0.71$\pm0.09$  & 78.2$\pm15.3$ & 69.9$\pm17.2$ & 0.75$\pm0.09$  & 82.3$\pm13.7$ & 68.5$\pm13.2$ & 0.77$\pm0.08$ \\ \hline
    100   & 62.4$\pm14.1$ & 80.6$\pm16.1$ & 0.68$\pm0.09$  & 81.3$\pm4.2$ & 66.2$\pm11.4$ & 0.76$\pm0.05$  & 79.6$\pm2.6$ & 70.5$\pm13.8$ & 0.76 $\pm0.05$ \\
    \bottomrule
    \end{tabular}%
  \label{tab:ResNet}%
  }
\end{table}
\begin{table}[t!]
{\scriptsize 
  \centering
  \setlength{\tabcolsep}{0.1em}
  \caption{Five-fold results for VGG16.}
    \begin{tabular}{c|c|c|c|c|c|c|c|c|c}
    \toprule
    \textbf{ \%} & \multicolumn{3}{c|}{\textbf{TL}} & \multicolumn{3}{c|}{\textbf{PRT+TL}} & \multicolumn{3}{c}{\textbf{All}} \\
\cmidrule{2-10}          & \textbf{SEN} & \textbf{SPE} & \textbf{F1} & \textbf{SEN} & \textbf{SPE} & \textbf{F1} & \textbf{SEN} & \textbf{SPE} & \textbf{F1} \\ \hline
    \midrule
    10    & 99.4$\pm1.3$ & 3.5$\pm4.8$  & 0.67$\pm0.01$  & 95.1$\pm6.3$ & 7.2$\pm6.2$  & 0.66$\pm0.02$  & 97.7$\pm4.3$ & 5.7$\pm7.9$  & 0.67$\pm0.01$\\ \hline
    25    & 94.8$\pm8.4$ & 18.0$\pm13.1$ & 0.69$\pm0.04$  & 87.4$\pm11.3$ & 30.6$\pm8.2$ & 0.68 $\pm0.04$ & 92.8$\pm6.2$ & 25.2$\pm4.2$ & 0.69$\pm0.03$ \\ \hline
    50    & 89.4$\pm13.9$ & 37.2$\pm23.4$ & 0.71$\pm0.04$  & 82.8$\pm12.3$ & 49.3$\pm6.5$ & 0.71$\pm0.06$  & 83.1$\pm13.1$ & 49.3$\pm6.3$ & 0.71$\pm0.06$ \\ \hline
    75    & 84.8$\pm19.3$ & 54.7$\pm22.3$ & 0.74$\pm0.06$  & 67.3$\pm7.1$ & 65.9$\pm16.2$ & 0.67$\pm0.04$  & 73.6$\pm5.3$ & 65.3$\pm10.7$ & 0.71$\pm0.05$ \\ \hline
    100   & 71.1$\pm13.1$ & 70.5$\pm19.7$ & 0.71$\pm0.05$  & 68.8$\pm14.7$ & 71.3$\pm6.9$ & 0.69$\pm0.08$  & 73.7$\pm13.1$ & 69.9$\pm4.3$ & 0.72$\pm0.08$ \\
    \bottomrule
    \end{tabular}%
    }
  \label{tab:VGG}%
\end{table}%
\begin{table}[t!]
{\scriptsize 
  \centering
  \setlength{\tabcolsep}{0.1em}
  \caption{Five-fold results for DenseNet201.}
    \begin{tabular}{c|c|c|c|c|c|c|c|c|c}
    \toprule
    \textbf{\%} & \multicolumn{3}{c|}{\textbf{TL}} & \multicolumn{3}{c|}{\textbf{PRT+TL}} & \multicolumn{3}{c}{\textbf{All}} \\
\cmidrule{2-10}          & \textbf{SEN} & \textbf{SPE} & \textbf{F1} & \textbf{SEN} & \textbf{SPE} & \textbf{F1} & \textbf{SEN} & \textbf{SPE} & \textbf{F1} \\
    \hline \midrule
    10    & 98.8$\pm1.2$ & 1.4$\pm1.8$  & 0.66$\pm0.01$  & 98.0$\pm2.4$ & 2.0$\pm2.2$  & 0.66$\pm0.01$  & 99.1$\pm0.8$ & 1.4$\pm1.4$ & 0.67$\pm0.01$ \\ \hline
    25    & 93.7$\pm1.6$ & 24.1$\pm5.5$ & 0.70$\pm0.01$  & 90.6$\pm8.4$ & 31.5$\pm6.9$ & 0.70$\pm0.04$  & 91.1$\pm10.5$ & 31.5$\pm9.9$ & 0.70$\pm0.03$ \\ \hline
    50    & 86.3$\pm6.2$ & 52.2$\pm10.0$ & 0.74$\pm0.06$  & 83.9$\pm7.1$ & 60.7$\pm10.5$ & 0.75$\pm0.07$  & 85.4$\pm6.3$ & 60.5$\pm11.7$ & 0.76$\pm0.07$ \\ \hline
    75    & 77.4$\pm7.3$ & 71.9$\pm14.1$ & 0.75$\pm0.07$  & 75.9$\pm11.0$ & 69.3$\pm6.9$ & 0.73$\pm0.09$  & 79.1$\pm9.6$ & 70.5$\pm10.9$ & 0.76$\pm0.09$ \\ \hline
    100   & 71.8$\pm8.6$ & 75.4$\pm9.6$ & 0.73$\pm0.08$  & 73.9$\pm5.2$ & 74.5$\pm9.4$ & 0.74 $\pm0.05$ & 76.2$\pm1.8$ & 75.4$\pm8.2$ & 0.76$\pm0.04$ \\
    \bottomrule
    \end{tabular}%
  \label{tab:DenseNet}%
}
\end{table}

\section{Conclusion}
Data scarcity and class imbalance are common problems faced in medical imaging and other practical domains. We proposed the concept of Pre-text Representation Transfer (PRT) that can mitigate the adverse effects of these problems. The PRT allows us to tap into the cheaply available unlabelled data of the domain. This unlabelled data is used to systematically transfer the representation component of the deep model to the target domain, without changing the classification component. This allows us to use potentially unlimited data in the transfer. This is in sharp contrast to the conventional Transfer Learning that can only use limited annotated data for model transfer. By applying this concept to CT-based COVID-19 detection task, we demonstrated that PRT can not only be used to construct a more effective feature extractor of the target domain, but it can also be used to boost the performance of the conventional Transfer Learning. Moreover, we also devised a mechanism to fuse the outputs of a PRT-enhanced model with a PRT-based feature extractor to further enhance the final performance. Our five-fold experiments with three large-scale visual models, using five data imbalance settings, thoroughly established the effectiveness of the proposed technique.
\section*{Acknowledgment}
This work was supported by Australian Government Research Training Program Scholarship. Dr.~Akhtar is  recipient of an Office of National
Intelligence  National Intelligence Postdoctoral Grant funded
by the Australian Government.

 \bibliographystyle{splncs04}
 \bibliography{paper}

\end{document}